\documentclass[aps,prl,twocolumn,noshowpacs,superscriptaddress,groupedaddress, floatfix]{revtex4}  
\usepackage{graphicx}  
\usepackage{dcolumn}   
\usepackage{bm}        
\usepackage{amssymb}   
\usepackage{amsmath}
\usepackage{color}
\usepackage[colorlinks=true, pdfstartview=FitV, linkcolor=blue, citecolor=blue, urlcolor=blue]{hyperref} 
\usepackage{epstopdf}
\usepackage[free-standing-units=true, per-mode=symbol-or-fraction, range-phrase={-}]{siunitx} 
\DeclareSIUnit{\sqrthz}{\ensuremath{\sqrt{\text{\hertz}}}} 

\usepackage{appendix} 

\newcommand{\affilIFP}{Laboratory for Solid State Physics, ETH Z\"{u}rich, CH-8093 Z\"urich, Switzerland.}
\newcommand{\affilAtto}{present address: attocube systems AG, DE-85540 Haar, Germany.}
\newcommand{\affilQuad}{present address: QUAD Systems AG, CH-8305 Dietlikon, Switzerland.}

\begin{document}

\preprint{}

\title{Vibrational Squeezing via Spin Inversion Pulses}

\author{Marc-Dominik Krass}\thanks{These authors contributed equally to this work.}
\affiliation{\affilIFP}
\affiliation{\affilAtto}
\author{Nils Prumbaum}
\thanks{These authors contributed equally to this work.}\affiliation{\affilIFP}
\author{Raphael Pachlatko}\affiliation{\affilIFP}\affiliation{\affilQuad}
\author{Christian L. Degen}\affiliation{\affilIFP}
\author{Alexander Eichler}\affiliation{\affilIFP}

\date{\today}

\begin{abstract}
Magnetic Resonance Force Microscopy (MRFM) describes a range of approaches to detect nuclear spins with mechanical sensors. MRFM has the potential to enable magnetic resonance imaging (MRI) with near-atomic spatial resolution, opening up exciting possibilities in solid state and biological research. In many cases, the spin-mechanics coupling in MRFM is engineered with the help of periodic radio-frequency pulses. In this paper, we report that such pulses can result in unwanted parametric amplification of the mechanical vibrations, causing misinterpretation of the measured signal. We show how the parametric effect can be cancelled by auxiliary radio-frequency pulses or by appropriate post-correction after careful calibration. Future MRFM measurements may even make use of the parametric amplification to reduce the impact of amplifier noise.
\end{abstract}


\maketitle


Nanomechanical sensors are excellent devices for spin detection and provide the basis for several ambitious proposals in quantum transduction and nanoscale imaging. On the one hand, spin-mechanics coupling is envisioned to enable readout and transfer of the polarization states of individual spins~\cite{Rabl_2010}. The realization of this proposal would allow quantum information exchange between remote spin qubits. On the other hand, spin-mechanics coupling also forms the basis of magnetic resonance force microscopy (MRFM)~\cite{Sidles_1991,Sidles_1992,rugar1992mechanical,poggio_force_2010}, which could become a transformative technology for nondestructive imaging of individual, complex biomolecules. While current proof-of-principle demonstrations are still too coarse-grained to reveal interesting structural information~\cite{Degen_2009,grob_magnetic_2019,Krass_2022}, the method will profit greatly from the progress achieved with optomechanical systems, and especially with high-$Q$ silicon nitride resonators~\cite{Tsaturyan_2017,Reetz_2019,Ghadimi_2018,Gisler_2022,Bereyhi_2022,Shin_2022,eichler2022ultra}.

Typical spin-mechanics experiments rely on a non-resonant, weak coupling between the spins in a sample and the mechanical sensor, mediated by a magnetic field gradient. Non-resonant coupling signifies that the resonance frequency $f_0$ of the sensor is much lower than the Larmor frequency $f_\mathrm{L} = \gamma B/2\pi$ of spins, where $\gamma$ is the gyromagnetic ratio and $B$ an applied magnetic field strength. In order to engineer efficient coupling between the spins and the sensor, a number of different protocols have been developed~\cite{Nichol_2013,mamin2007nuclear,vinante2011magnetic,Kosata_2020}. A commonly used method relies on pulsed radio-frequency (rf) magnetic fields to periodically invert the spins~\cite{Degen_2009,grob_magnetic_2019}. With a pulse repetition rate of $2f_0$, the interaction between the spins and a magnetic field gradient generates a force at $f_0$ that drives the sensor into measurable oscillations.

Most MRFM setups operate in the weak-coupling regime, where the averaging time required to pick up a spin signal is much longer than the effective spin lifetime in the rotating frame $\tau_\mathrm{m}$~\cite{Degen_2007,Degen_2008,slichter2013principles}. In addition, the thermal spin polarization is negligible for small spin ensembles. As a consequence, the measured mechanical oscillation does not reflect the instantaneous spin ensemble polarization. Instead, the stochastic fluctuations of the spin ensemble over times $t \gg \tau_\mathrm{m}$ lead to a force noise that increases the variance of the sensor's oscillation~\cite{Degen_2007,herzog2014boundary}. By selecting the phase of the spin inversion pulses, the phase of the increased variance can be controlled. The resulting sensor fluctuations in phase space still have a Gaussian distribution in both $X$ and $Y$, but one of the quadratures shows an increase in the variance.

In this paper, we reveal that the pulsed spin inversion method can produce a spurious driving effect that manifests as an increase in the sensor oscillation variance in one quadrature. This effect, while observed and heuristically avoided in the past, is little understood. The spurious driving closely resembles a real spin signal and can therefore lead to misinterpretation of data. We propose that the observed effect is due to phase-dependent parametric amplification (squeezing) of the sensor's thermomechanical fluctuations. We demonstrate that the squeezing artifact can be suppressed by the addition of a second set of pulses between the spin inversion pulses, which ``unsqueezes'' the phase space distribution. In this way, we are able to obtain an artifact-free spin signal.

\begin{figure*}[htb]
\includegraphics[width=\textwidth]{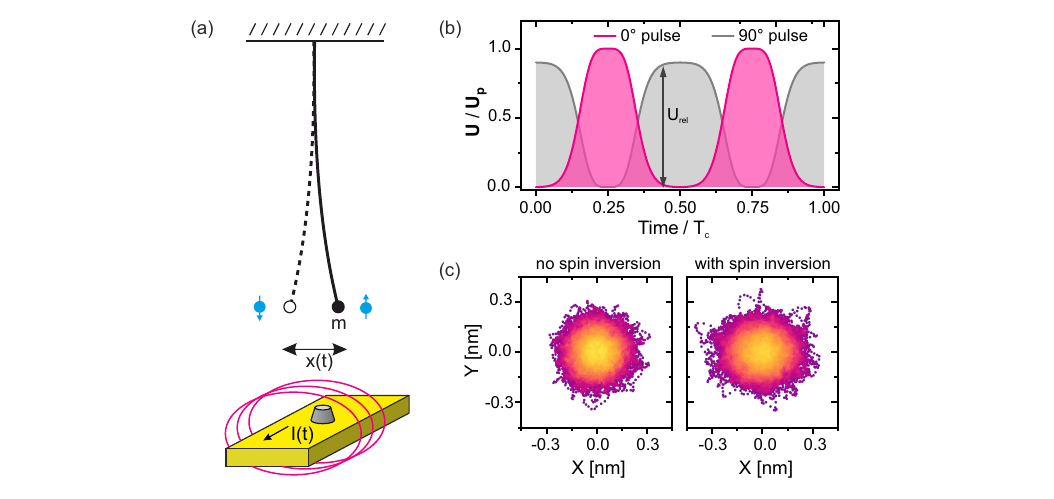}
\caption{Nanomechanical spin detection protocol. (a)~Illustration of the MRFM setup comprising a spin ensemble (blue arrow) at the tip of a cantilever sensor with mass $m$. A pulsed current $I(t)$ through a microstrip produces a magnetic field (pink lines) that periodically inverts the spin ensemble polarization $I_\mathrm{z}$. The interaction between $I_\mathrm{z}$ and a magnetic field gradient from a nanomagnet (grey cone) creates a force that drives cantilever oscillations $x(t)$. (b)~Amplitude modulation of voltage $U$ (with peak voltage $U_p$) applied for spin inversion pulses ($\SI{0}{\degree}$) and off-resonant fill pulses ($\SI{90}{\degree}$) over one cantilever period $T_\mathrm{c}$. (c)~Measured fluctuations of the cantilever oscillations in a phase space rotating at $f_0$ without and with spin inversion pulses. The color code indicates probability density from low (pink) to high (yellow). The fluctuating spin force manifests as an increased variance $\sigma_\mathrm{X}^2$.}
\label{fig:fig1}
\end{figure*}


In our setup, the fundamental mode of a silicon cantilever acts as the mechanical sensor. The cantilever is positioned in the pendulum geometry above a gold microstrip fabricated on top of a thermally oxidized silicon chip. The cantilever has a resonance frequency $f_0 = \SI{3500}{\hertz}$, an effective mass $m = \SI{e-13}{\kilo\gram}$, and a quality factor $Q = \num{25000}$. For spin-mechanics experiments, a sample is attached to the tip of the cantilever and colled down to $T\approx\SI{5}{\kelvin}$. The spin ensemble inside the sample, which is the typical subject of study in MRFM, is illustrated by a single blue spin in Fig.~\ref{fig:fig1}(a).

In order to manipulate the spin ensemble, amplitude- and frequency-modulated rf current pulses with a carrier frequency around $f_\mathrm{L}$ are sent through the microstrip on the chip surface~\cite{grob_magnetic_2019}. The current generates rf magnetic fields that flip spins once every pulse, see pink outlines in Fig.~\ref{fig:fig1}(b). With a pulse repetition rate of $2f_0$, the interaction of the $z$-component of the spin ensemble $I_\mathrm{z}$ with the magnetic field gradient of a nanoscale ferromagnet [Fig. 1(a)] creates a periodic force that drives the cantilever oscillation at $f_0$. Stochastic spin fluctuations with lifetime $\tau_\mathrm{m}$ slowly change $I_\mathrm{z}$ and average the mean force signal to zero over long integration times $t\gg \tau_\mathrm{m}$. For this reason, it is usually the added oscillation variance $\sigma_\mathrm{spin}^2$ caused by the fluctuating spin force that serves as the spin signal in MRFM~\cite{Degen_2007,herzog2014boundary}. By selecting the pulse phase relative to the lock-in amplifier clock at $f_0$, the phase of $\sigma_\mathrm{spin}^2$ can be controlled; in the example shown in Fig.~\ref{fig:fig1}(c), the spin signal is chosen to be in the X channel. The spin force manifests as a difference between the variances in the two quadratures, $\sigma_\mathrm{spin}^2 = \sigma_\mathrm{X}^2-\sigma_\mathrm{Y}^2$. Note that the pulses at $2f_0$ do not cause direct electrostatic driving of the cantilever mode at $f_0$ because they do not break the symmetry over one period $T_\mathrm{c} = 1/f_0$.

\begin{figure}[t]
\includegraphics[width=\columnwidth]{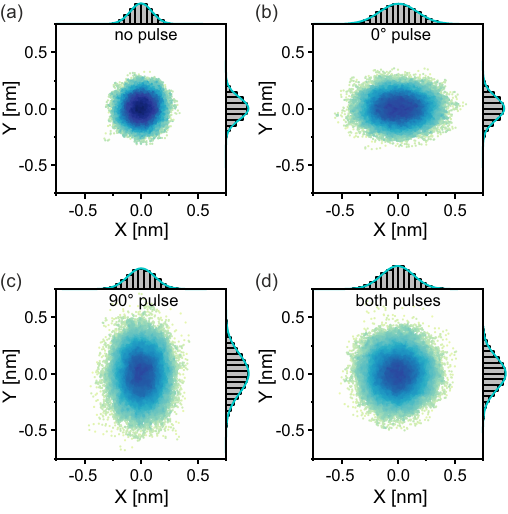}
\caption{Fluctuations measured in the absence of spin inversions with detuned pulses. The color code indicates probability density from low (yellow) to high (dark blue). Histograms quantify the distributions in $X$ and $Y$ in phase space. (a)~Thermomechanical noise of the cantilever mode when no pulses are applied. (b)~Fluctuations in the presence of \SI{0}{\degree} pulses. (c)~Fluctuations in the presence of \SI{90}{\degree} pulses. (d)~Fluctuations in the presence of both \SI{0}{\degree} and \SI{90}{\degree} pulses.}
\label{fig:fig2}
\end{figure}

Surprisingly, a significant imbalance between $\sigma_\mathrm{X}^2$ and $\sigma_\mathrm{Y}^2$ can be observed experimentally even when the pulses are detuned from $f_\mathrm{L}$ and do not excite any spins. In such a situation, one would expect that the pulses have no effect on the cantilever mode and that the phase space portrait of the thermal fluctuations remains circular as in Fig.~\ref{fig:fig2}(a). Instead, we clearly observe a significant imbalance $\sigma_\mathrm{X}^2 > \sigma_\mathrm{Y}^2$ in Fig.~\ref{fig:fig2}(b). This imbalance could be misinterpreted as a spin signal. In the past, this spurious driving was carefully avoided by heuristic pulse optimization~\cite{longenecker2012high,moores2015accelerated,grob_magnetic_2019,Krass_2022,pachlatko2024nanoscale}. When the phase of the pulse is rotated by \SI{90}{\degree}, the resulting variance is rotated as well, yielding $\sigma_\mathrm{Y}^2 > \sigma_\mathrm{X}^2$ in Fig.~\ref{fig:fig2}(c). When combining both sets of pulses, we return to a balanced distribution $\sigma_\mathrm{X}^2 \approx \sigma_\mathrm{Y}^2$, see Fig.~\ref{fig:fig2}(d). Here, both quadratures are slightly enlarged relative to Fig.~\ref{fig:fig2}(a), indicating an increase in the effective cantilever mode temperature. 

\begin{figure}[t]
\includegraphics[width=\columnwidth]{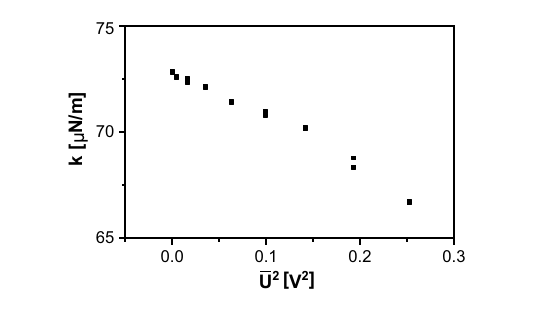}
\caption{Spring constant ($k = 4\pi^2m f_0^2$) of the cantilever mode measured as a function of average pulse power $\bar{U}^2$. We observe a roughly linear decrease of $k$ with $\bar{U}^2$.}
\label{fig:fig3}
\end{figure}

To understand the observations in Fig.~\ref{fig:fig2}, we need to consider two independent effects. On the one hand, current pulses dissipate energy, heating the cantilever mode irrespective of the pulse shape or phase. We assign the increase of $\sigma_\mathrm{X}^2$ and $\sigma_\mathrm{Y}^2$ in Fig.~\ref{fig:fig2}(d) relative to Fig.~\ref{fig:fig2}(a) to such Joule heating. On the other hand, we observe
that the squared field strength associated with the spin inversion pulses can modify the cantilever spring constant, see Fig.~\ref{fig:fig3}. We tentatively associate this effect with electrostatic interactions between the biased surface and random charges on the cantilever tip~\cite{heritier2021spatial}. When the field power is modulated in time with a rate close to $2f_0$, it causes positive and negative parametric amplification of the orthogonal oscillation quadratures~\cite{Rugar_1991,Lifshitz_Cross,Eichler_Zilberberg_book}.

\begin{figure}[t]
\includegraphics[width=\columnwidth]{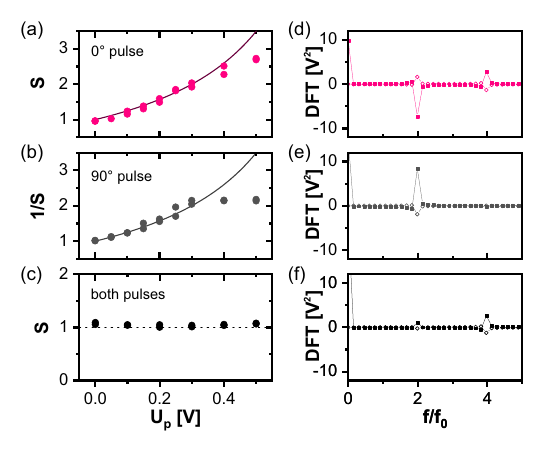}
\caption{(a)~Squeezing factor $S = \sigma_\mathrm{X}^2 / \sigma_\mathrm{Y}^2$ as a function of the pulse amplitude $U_\mathrm{p}$ for \SI{0}{\degree} pulses applied to the cantilever. A line quantifies the expected trend, see main text. This trend describes the data well up to $U_\mathrm{p} \approx \SI{0.4}{\volt}$, beyond which we found instabilities in our pulse generation setup. (b)~Same as in (a) but for \SI{90}{\degree} pulses, showing $1/S$ instead of $S$. (c)~Squeezing factor when both pulses are applied simultaneously. A dashed line at $S = 1$ is a guide to the eye. (d)-(f)~Discrete Fourier transforms (DFT) of the measured pulse shapes used in (a)-(c), respectively. Real and imaginary components of the DFT are shown as filled squares and open spheres, respectively. We find that the DFTs have positive, negative, and vanishing amplitudes at $f/f_0 = \pm 2$ for the three respective situations.}
\label{fig:fig4}
\end{figure}

To show that parametric amplification can be used to model our experimental observations, we examine the measured squeezing factor $S = \sigma_\mathrm{X}^2 / \sigma_\mathrm{Y}^2$ in Fig.~\ref{fig:fig4}. When only the `\SI{0}{\degree}' rf pulses are applied (without inverting any spins), $S$ increases monotonically with the maximum pulse amplitude $U_\mathrm{p}$. By contrast, when only the rotated `\SI{90}{\degree}' pulses are used, the inverse $1/S$ increases monotonically with $U_\mathrm{p}$. Both findings are in agreement with the observations in Fig.~\ref{fig:fig2}. Beyond $U_\mathrm{p}\approx \SI{0.4}{\volt}$, the squeezing saturates for both pulse types. In this voltage regime, we found spurious effects in our pulse protocol that may cause further artifacts. We avoid this voltage range in our MRFM experiments and also ignore it in the following discussion.

To quantify the changes in $S$, we plot in Fig.~\ref{fig:fig4}(a) and (b) the expected parametric squeezing ratio $(1+\beta U_\mathrm{p})/(1-\beta U_\mathrm{p})$~\cite{Lifshitz_Cross} as solid lines, where $\beta = \SI{1.11}{\volt^{-1}}$ is a heuristic factor to account for the interaction efficiency between the pulses and the cantilever displacement. This simple model accounts well for the observed increase in $S$ and $1/S$, respectively, in the relevant range $U_\mathrm{p} < \SI{0.4}{\volt}$. When parametric amplification is applied to both quadratures simultaneously, symmetry between fluctuations in $X$ and $Y$ should be restored. Indeed, in Fig.~\ref{fig:fig4}(c) we show that $S\approx 1$ when both \SI{0}{\degree} and \SI{90}{\degree} are combined. This entails that \SI{90}{\degree} pulses can be used to counter unwanted squeezing during spin detection measurements.

The origin of the parametric squeezing, and its cancellation by the combination of \SI{0}{\degree} and \SI{90}{\degree} pulses, can be confirmed by a Fourier analysis of the applied pulse shapes. In Fig.~\ref{fig:fig4}(d), we display the discrete Fourier transform (DFT) of the measured squared pulse voltage (the pulse power) for the \SI{0}{\degree} pulse. The spectrum has a peak at $2f_0$, as expected from the amplitude modulation of the pulse, as depicted in Fig.~\ref{fig:fig1}(b). This Fourier component at $2f_0$ is responsible for parametric amplification and squeezing of the cantilever oscillations. We obtain the same result for the \SI{90}{\degree} pulse in Fig.~\ref{fig:fig4}(e). However, the sign of the component at $2f_0$ is inverted, as expected for the DFT of a squared and $\pi/2$ phase-shifted sinusoidal signal. Finally, when both pulses are combined, the positive and negative peaks of the pulses cancel and the resulting spectrum has almost no signature near $2f_0$. As a consequence, no parametric squeezing effects are present.

Note that the \SI{0}{\degree} and \SI{90}{\degree} pulses can have different amplitude modulation functions and peak amplitudes, c.f. Fig.~\ref{fig:fig1}(b). As long as the $2f_0$-component of both pulses is equal in magnitude, the parametric squeezing is compensated. This enables significant freedom in optimizing spin inversion protocols.

In summary, we reveal that spin inversion pulses in MRFM can result in parametric squeezing of cantilever vibrations, which yields a signal that closely resembles that of a real spin force. The effect can be cancelled by combining two sets of phase-shifted pulses: the \SI{0}{\degree} pulses are applied at a carrier frequency $f_\mathrm{L}$ to invert nuclear spins within a selected Larmor frequency band, while the \SI{90}{\degree} pulse are detuned from $f_\mathrm{L}$ and do not excite spins. This method is very robust: once a suitable \SI{90}{\degree} pulse is found, the compensation works regardless of the instantaneous cantilever frequency or the pulse amplitude scaling, which is very beneficial for scanning experiments. A disadvantage of adding the \SI{90}{\degree} pulses is increased Joule heating, as shown in Fig.~\ref{fig:fig2}(d). For this reason, it is worth considering alternative methods for reducing the squeezing effect of the \SI{0}{\degree} pulses.

With a careful calibration of the parametric interaction, the squeezing can be removed from the collected spin force data in post analysis by applying the function inverse $(1-\beta U_\mathrm{p})/(1+\beta U_\mathrm{p})$, where the value of $\beta$ can be obtained from a measurement series such as shown in Fig.~\ref{fig:fig4}(a). With this method, no second pulse is required to cancel the resonator oscillation squeezing, and hence Joule heating is reduced. Squeezing can potentially even turn into a resource for enhancing the spin signal relative to amplifier noise, leading to an enhanced signal-to-noise ratio. However, note that the ratio between the measured spin force and force fluctuations acting on the sensor is not changed by squeezing, hence no sensitivity increase relative to the dominant thermomechanical force noise is expected.

We expect that the understanding of parametric effects related to spin driving will enable researchers to design better pulse shapes via optimal control theory~\cite{rose_high_2018} and machine learning, thereby leading to improved spin sensing protocols. Such design rules will be crucial for establishing spin sensing protocols with mechanical sensors in the MHz regime. These are expected to improve spin sensitivity, but come with the need for much faster nuclear spin manipulations~\cite{Kosata_2020,eichler2022ultra,haas2022nuclear,tabatabaei2024large}.

\section*{Acknowledgments}
We thank the operations team of the FIRST cleanroom, especially Sandro Loosli and Petra Burkard, as well as the operations team of the Binnig and Rohrer Nanotechnology Center (BRNC) at IBM Rüschlikon, especially Ute Drechsler, Richard Stutz and Dr. Diana Dávila Pineda. A. E. acknowledges financial support from Swiss National Science Foundation (SNSF) through grants 200021\_200412 and CRSII5\_206008/1. 

\vspace{5mm}
The authors have no conflicts to disclose.

The data that support the findings of this study are available from the corresponding author upon reasonable request.

\bibliographystyle{apsrev4-1}
\bibliography{references}

\end{document}